\documentclass{article}

\usepackage{graphicx}

\def\And{{\rm and\ }}

\def\stars{\bigskip\centerline{***}\medskip}

\newif\ifboo \boofalse

\def\Review#1{\boofalse{\it #1},}
\def\Name#1{{\sc #1},}
\def\Vol#1{\ifboo Vol. {\bf #1}\else{\bf #1}\fi}
\def\Year#1{\ifboo #1\else(#1)\fi}
\def\Book#1{\bootrue{\it #1},}
\def\Page#1{\ifboo {\rm p. #1}\else{\rm #1}\fi}

\begin{document}

\title{\bf On the center of mass of Ising vectors}

\author{M. Copelli, M. Bouten, C. Van den Broeck \And B. Van Rompaey 
\\ \ \\ Limburgs Universitair Centrum \\ B-3590 Diepenbeek \\ Belgium}

\date{}


\maketitle

\begin{flushleft}
statistics 
mechanics of model systems 
\end{flushleft}

\begin{abstract}
We show that the center of mass of Ising vectors that obey some simple 
constraints, is again an Ising vector.
\end{abstract}


Many problems in statistical mechanics are formulated in terms of an 
N-dimensional vector ${\bf J}$, with components $J_i, i=1,...,N$ that 
take only binary values $J_i=\pm 1$.  We will call such a vector an 
Ising vector.  Its components represent for example a spin state 
(Ising model \cite{ising}), the occupancy of a site etc.  In the 
thermodynamic limit $N \rightarrow \infty$, only a subset of all the 
possible configurations $\{{\bf J}\}$ are typically realized.  In many 
cases, they are characterized by simple contraints of the form :

\begin{equation}\label{c}
\lim_{N\to\infty}\frac{{\bf J}\cdot{\bf B}}{N}= R\;\;\;\;\;\;\;\;\; 
\lim_{N\to\infty}\frac{{\bf J}\cdot{\bf J}^{\prime}}{N}=q,
\end{equation}
where ${\bf J}$ and ${\bf J}^{\prime}$ are typical members of the 
subset, ${\bf B}$ is a symmetry breaking direction (imposed from the 
outside or arising through a phase transition), while $q$ and $R$ are 
physical properties describing the resulting macroscopic state (for 
example the magnetization or the density).  In this letter, we focus 
on the center of mass of the vectors ${\bf J}$, that satisfy the above 
constraints.  We report the surprising finding that it is an Ising 
vector whenever ${\bf B}$ is Ising.

To construct the center of mass we follow a Monte Carlo approach by 
choosing at random $n$ vectors ${\bf J}^a, a=1,...,n$, that satisfy 
the constraints, and considering their center of mass:
\begin{equation}
	{\bf Y} =C^{-1} \;\sum_{a}
	{\bf J}^{a},
\end{equation}
with the proportionality factor $C=\sqrt{n+n(n-1)q}$, so that the 
normalization condition ${\bf Y}^2=N$ is obeyed.  In general the 
vector ${\bf Y}$ is not Ising, but our contention is that it becomes 
Ising in the limit $n \rightarrow \infty$, provided ${\bf B}$ is 
Ising.  In view of the permutation symmetry between the coordinate 
axes, it will be sufficient to prove that $B_1 
Y_1=C^{-1}\sum_{a}B_{1}\;J^{a}_{1}$ only takes the values $+1$ and 
$-1$ in this limit.  To show that this is the case, we focus our 
attention on the probability density $P(y)$ of the variable 
$y=n^{-1}\sum_{a}B_{1}\;J^{a}_{1}$, which differs by a factor 
$n^{-1}\sqrt{n+n(n-1)q}\stackrel{n\to\infty}{\to}\sqrt{q}$ from 
$B_{1}Y_{1}$.  It is given by:
\begin{eqnarray}
	\label{py}
	P(y) & \sim &\int
	\left[\prod_{a}^{n}d{\bf  J}^{a}P_{b}({\bf  J}^{a})
	\;\delta({\bf  J}^{a}\cdot{\bf B}-NR)\right] \nonumber \\
 & \times & \left[\prod_{a<b}\delta({\bf  J}^{a}\cdot{\bf  J}^{b}-Nq)\right]
	\delta\left(y - n^{-1}\sum_{a}B_{1}\;J^{a}_{1}\right)\; ,
\end{eqnarray}
where $P_{b}$ is the measure restricting to vectors with binary
components,
\begin{equation}
	\label{pb}
	P_{b}({\bf  J}) = \prod_{j=1}^{N}\left[
	\frac{1}{2}\delta(J_{j}-1) + \frac{1}{2}\delta(J_{j}+1)
	\right] \;,
\end{equation}
and the proportionality constant has to be determined from the 
normalization condition $\int_{-\infty}^{\infty}\;dy\;P(y)\;=1$.  The 
r.h.s. of (\ref{py}) resembles an ordinary replica calculation 
\cite{mezard}, but with as limit of interest the number of replicas 
$n$ tending to infinity.

\begin{figure}[h,t]
\begin{center}
\includegraphics[width = \textwidth]{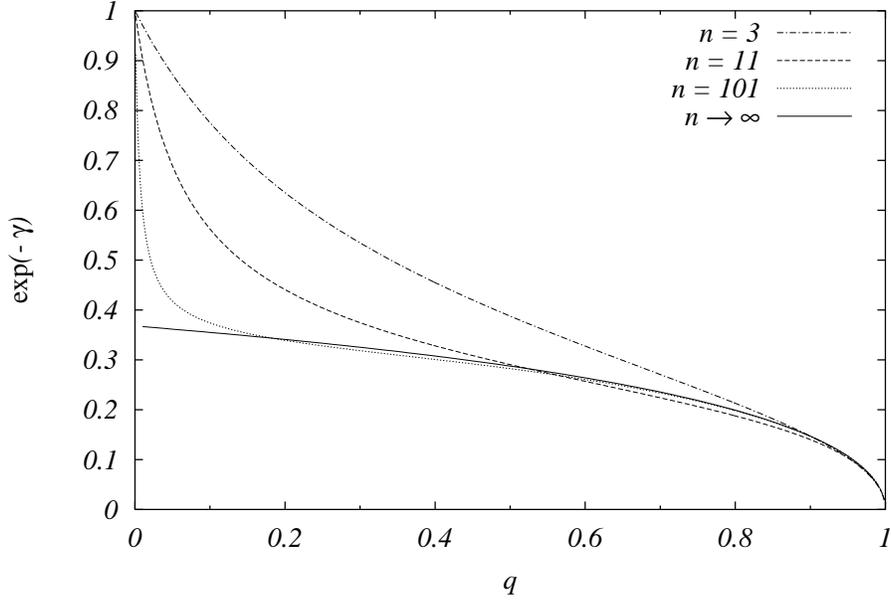}
\end{center}
\caption[short]{\label{fig:gamma}
In order to account for a logarithmic divergence in the limit $q\to 
1$, $\exp(-\gamma),\gamma = n\hat{q},$ is plotted as a function of $q$ 
for several values of $n$.  
}
\end{figure}

\begin{figure}[h,b]
\begin{center}
\includegraphics[width = \textwidth]{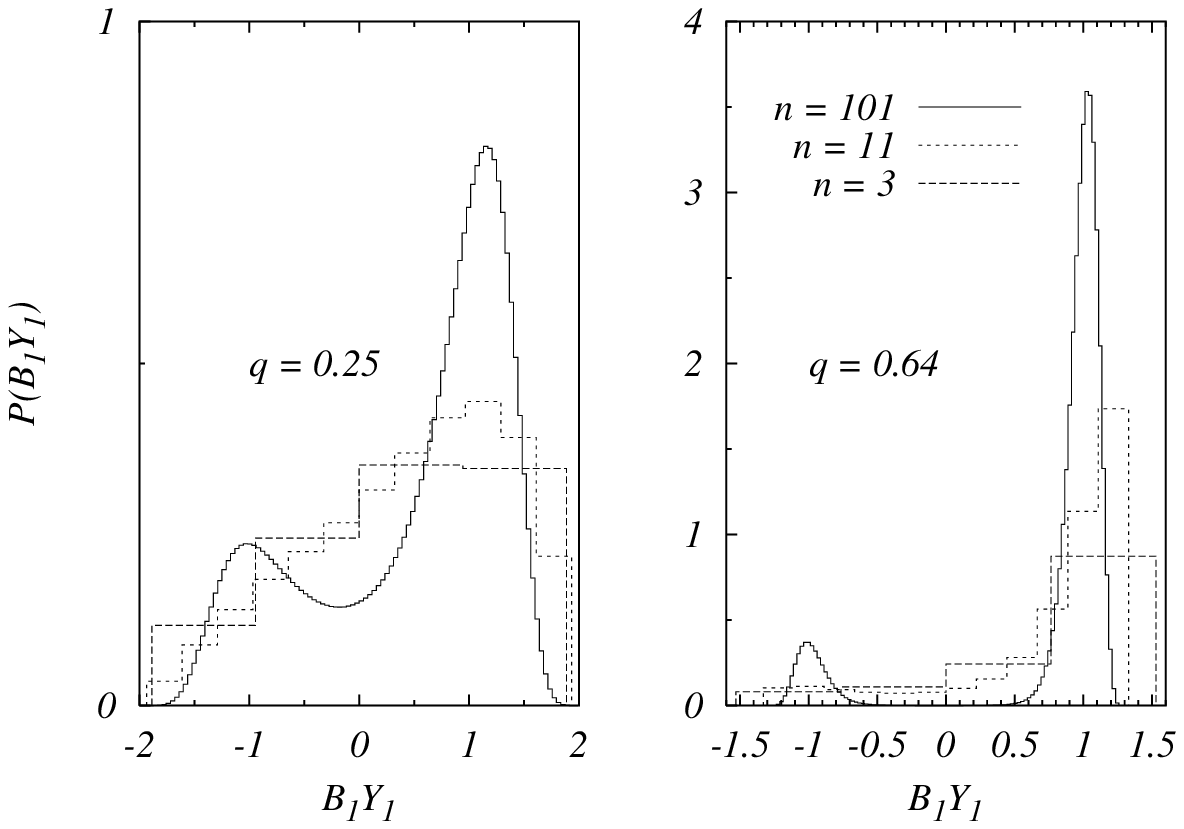}
\end{center}
\caption[short]{\label{fig:ProbGibbs}
Probability density for $B_{1}Y_{1}$ according to eq.~(\ref{pyme}) for 
$q=R$ and several values of $n$.  The legend is the same for both 
plots.}
\end{figure}

Rather than following the standard but lengthy calculations that are 
usual in this case, we present a more elegant, direct and expedient 
procedure.  Since $y=n^{-1}\sum_a x_a$, with $ x_{a}= J^{a}_{1}B_{1}$, 
we evaluate the joint probability density $P({\bf x})$ of the 
$n$-dimensional vector with binary components $x_a$, $ a=1,\ldots,n $.  
Since all choices of the vectors ${\bf J}^{a}$ that satisfy the 
constraints can be realized, the Shannon entropy is maximized under 
the constraints (\ref{c}).  Hence $P({\bf x})$ is found by 
maximizing~\cite{Jaynes} its Shannon entropy $-\sum_{\bf x}P({\bf 
x})\ln P({\bf x})$, subject to the constraints:
\begin{eqnarray}
	\label{constraints}
	\left<x_{a}\right> & = & \frac{{\bf J}^{a}\cdot {\bf B}}{N} = R
	  \nonumber \\
	\left<x_{a}x_{b}\right> & = &
	\frac{{\bf J}^{a}\cdot{\bf J}^{b}}{N} = q  \ \ \ \ \ \ (a<b).
\end{eqnarray}
One finds:
\begin{equation}
	\label{pxa}
	P({\bf x}) = {Z^{-1}}
	\exp\left[
	\sum_{a}\hat{R}_{a}x_{a} +\sum_{a<b}\hat{q}_{ab}x_{a}x_{b}
	\right]\; ,
\end{equation}
where $Z$ follows from the normalization of $ P({\bf x})$.  The values 
of the Lagrange multipliers $\{\hat{R}_{a}\}$ and $\{\hat{q}_{ab}\}$ 
have to determined from the constraints (\ref{constraints}).  In view 
of the permutation symmetry in the replica indices, $\hat{R}_{a}$ and 
$\hat{q}_{ab}$ must be independent of $a$ and $b$, $\hat{R}_{a} = 
\hat{R},\; \hat{q}_{ab} = \hat{q}$, rendering the evaluation of $Z$ 
very simple:
\begin{equation}
	\label{ZMERS}
	Z(\hat{R},\hat{q}) = \; e^{-n\hat{q}/2}\int Dz\left[
	\cosh\left(\hat{R} + z\sqrt{\hat{q}}\right)
	\right]^{n}\; ,
\end{equation}
while  (\ref{constraints}), determining
$\hat{R}$ and
	$\hat{q}$, reduce to:
\begin{eqnarray}
	\label{sp0}
	R & = & \frac{1}{n}\;\frac{\partial}{\partial\hat{R}}\ln Z
 =
	\frac{\int du\,\exp\left[ -(u-\hat{R})^{2}/2\hat{q} \right]
	(\cosh u)^{n}\,\tanh u}
	{\int du\,\exp\left[ -(u-\hat{R})^{2}/2\hat{q} \right]
	(\cosh u)^{n}}
	\nonumber \\
	q & = & \frac{2}{n(n-1)}\;\frac{\partial}{\partial\hat{q}}\ln Z\
  =
	\frac{\int du\,\exp\left[ -(u-\hat{R})^{2}/2\hat{q} \right]
	(\cosh u)^{n}\,\tanh^{2} u}
	{\int du\,\exp\left[ -(u-\hat{R})^{2}/2\hat{q} \right]
	(\cosh u)^{n}}\; .
\end{eqnarray}
As a result of the ``replica symmetry'', we conclude from (\ref{pxa}) 
that $P({\bf x})$ is in fact a function of $\sum_a \; x_a = n y$.  
Hence $P(y)$ is obtained from $P({\bf x})$ by multiplication with a 
combinatorial factor, expressing the freedom to choose which 
$n(1+y)/2$ components of ${\bf x}$ are $+1$ and which remaining ones 
are $-1$, for a given total value $ny$ of their sum:
\begin{equation}
	\label{pyme}
	P(y) \sim \left(
	\begin{array}{c}
		n \\ \frac{n(1+y)}{2}
	\end{array}
	\right)\;{\exp\left[
	n\hat{R} y + \frac{n^{2}\hat{q} y^{2}}{2}\right]}
	\; .
\end{equation}
$y$ can take the values $-1,-1+2/n,\ldots,1-2/n,1$, and the 
proportionality constant is again fixed by normalization.  This result 
is in agreement with a direct evaluation of~(\ref{py}) but very 
different from the result for continuous components discussed 
in~\cite{Schietse95}.  Unfortunately, the above expression is quite 
complicated, especially in view of the fact that we did not succeed in 
solving explicitly the eqs.~(\ref{sp0}) determining the Lagrange 
multipliers.  Concordantly, the components of ${\bf Y}$ are not binary 
for any finite $n$.  As an illustration, we have included in 
fig.~\ref{fig:gamma} the results obtained by a numerical solution 
of~(\ref{sp0}) for the special case $q=R$ and several values of $n$.  
The corresponding results for the probability density for $y$ (or 
equivalently, $B_{1}Y_{1}$), are plotted in fig.~\ref{fig:ProbGibbs}.

In order to extract the asymptotic behavior for the $n \rightarrow 
\infty$ limit, one needs to guess the asymptotic dependence on $n$ of 
the Lagrange parameters.  The correct scaling appears quite naturally 
in the calculations for the simpler case of vectors ${\bf J}$ with 
continuous components.  Here, we just note by inspection that the 
eqs.~(\ref{sp0}) for the properly scaled Lagrange parameters $\rho 
\equiv n\,\hat{R}$ and $\gamma \equiv n\,\hat{q}$ read:
\begin{eqnarray}
	\label{sp2}
	R & = &
	\frac{\int du\, e^{-n\phi(u)}\sinh(u\rho/\gamma)\tanh u}
	{\int du\, e^{-n\phi(u)}\cosh(u\rho/\gamma)}\nonumber \\
	q & = &
	\frac{\int du\, e^{-n\phi(u)}\cosh(u\rho/\gamma)\tanh^2 u}
	{\int du\, e^{-n\phi(u)}\cosh(u\rho/\gamma)}\; ,
\end{eqnarray}
where
\begin{equation}
	\label{phi}
	\phi(u) \equiv \frac{u^{2}}{2\gamma} - \ln\cosh u \; .
\end{equation}
The appearance of the hyperbolic functions of $u\rho/\gamma$ in 
eqs.~(\ref{sp2}) is due to the fact that $\phi$ is even.  The saddle 
point approximation can now, for $n \rightarrow \infty$, be applied in 
a straightforward manner on the $u$-integrations, leading to the 
following simple and explicit solutions for the scaled Lagrange 
variables:
\begin{eqnarray}
	\label{spresults}
	\gamma & = &
	\frac{\mbox{arctanh}\sqrt{q}}{\sqrt{q}} \nonumber \\
	\rho & = & \frac{\mbox{arctanh}(R/\sqrt{q})}{\sqrt{q}}\; .
\end{eqnarray}
Inserting this result together with the asymptotic expression for the 
combinatorial factor in (\ref{pyme}), one finally obtains the 
following asymptotic result for $P(y)$:
\begin{eqnarray}
	\label{pyresult}
	P(y) &{\sim} &
	{\exp({\rho \;y})\exp n\left[
	\gamma \; y^{2}/2 - \ln\sqrt{1-y^{2}} - y\,\mbox{arctanh}\,y
	\right]}
	 \nonumber
	\\ & \stackrel{n\to\infty}{\to} &
	\frac{1}{2}\left( 1+ \frac{R}{\sqrt{q}} \right) \delta(y-\sqrt{q}) +
	\frac{1}{2}\left( 1- \frac{R}{\sqrt{q}} \right) \delta(y+\sqrt{q})\; .
\end{eqnarray}
In view of the aforementioned relation between $y$ and the (first) 
component of ${\bf Y}$, the convergence of the latter to an Ising 
vector follows immediately.

As a first application of the above result, we turn to the case of an 
Ising spin system in the ferromagnetic phase.  Choosing the vector 
${\bf B}$ with all its components equal to $1$, we note that $R$ plays 
the role of the magnetization.  All the spin states ${\bf J}$ are 
otherwise allowed, and lie on the rim of the N dimensional sphere with 
radius $\sqrt{N}$ at fixed angle $\arccos R$ with ${\bf B}$.  It is 
thus clear that the center of mass is ${\bf B}$ itself.  This trivial 
result is recovered from (\ref{pyresult}) by noting that $q=R^2$ in 
this case.  Note that the constraint ${\bf J}^{a}\cdot{\bf J}^{b}=N q$ 
is therefore redundant, which implies $\hat{q}=0$ and 
$\hat{R}=\mbox{arctanh}R$ (a result valid for any $n$).

A case of special symmetry is $q = R$.  In this scenario, no 
macroscopic measure allows to distinguish between the symmetry 
breaking direction ${\bf B}$ and each of the vectors ${\bf J}^{a}$.  
The Lagrange parameters also present the symmetry $\hat{R} = \hat{q}$, 
which can be seen from eqs.~(\ref{sp0}).

A third case of interest is the limit $R \rightarrow 0$ while $q$ 
remains finite.  In this case, the ${\bf J}$-vectors lie in the 
subspace orthogonal to ${\bf B}$ and satisfy as single constraint the 
prescribed mutual overlap $q$.  From eqs.~(\ref{sp0}) it is clear that 
$\hat{R}=0$ is automatically satisfied, and one concludes from 
(\ref{pyresult}) that the center of mass of the ${\bf J}$-vectors is 
again an Ising vector (but its components are equally likely to be 
$+1$ or $-1$).

\begin{figure}[h,t]
\begin{center}
\includegraphics[width = \textwidth]{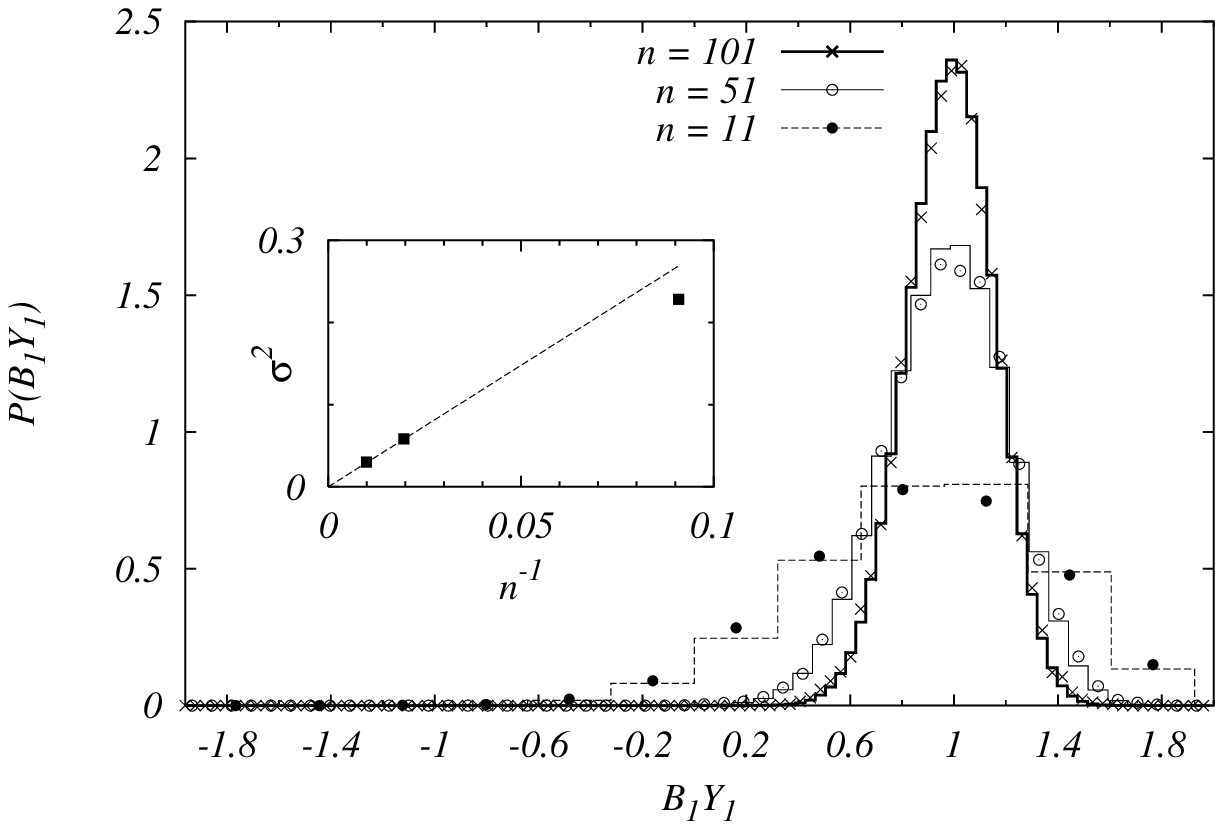}
\end{center}
\caption[short]{\label{fig:ProbIsing}
Probability density for $B_{1}Y_{1}$ in the $q=R^{2}$ scenario and 
several values of $n$.  The lines represent the theoretical 
curve~(\ref{pyme}), while points represent results from a simulation 
of the mean field ferromagnetic Ising model with $N=100$, 
dimensionless temperature 1.09 and dimensionless magnetic field 0.1 
(amounting to a magnetization $R\simeq 0.5$ -- see text for details).  
Inset: second cumulant ($\sigma^{2}$) of the distribution as a 
function of $1/n$.  The squares represent the simulations, while the 
dashed straight line corresponds to the theory (in the asymptotic 
limit $n\to\infty$, eq.~(\ref{pyresult}) implies 
$\sigma^{2}\sim (1-R^{2})/(nR^{2})$).}
\end{figure}

It is very tempting to apply the results above to neural network 
learning problems~\cite{hertz} where a student perceptron ${\bf J}$ 
learns from examples generated by a teacher perceptron ${\bf 
B}$~\cite{gardner,tishby,seung}.  Indeed, so-called Gibbs 
learning~\cite{gyorgyi} presents the symmetry $q=R$, and the interest 
of the center of mass is that it is, according to a simple general 
argument~\cite{watkin}, see also~\cite{opper}, the ``best'' student 
having the largest overlap with the teacher (namely $\sqrt{R}$).  
Accordingly, $R = 0$ and $q\neq 0$ are constraints satisfied by Ising 
vectors which solve the capacity problem~\cite{storage}.  However, 
these are disordered systems, and the conditions on $R$ and $q$ alone 
do not convey all the information which is necessary to describe the 
constraints in ${\bf J}$ space.  Therefore, even though the 
constraints~(\ref{c}) are satisfied in neural network problems, 
result~(\ref{pyresult}) does not apply to them.

We conclude with a verification of the theoretical prediction 
(\ref{pyme}) by running simulations for the mean field ferromagnetic 
Ising spin model, fig.~\ref{fig:ProbIsing}.  The Metropolis algorithm 
was allowed to run for a number of Monte Carlo steps per site (MCS) 
until thermalization was considered to be achieved.  Then vectors were 
sampled every 5 MCS (to allow sufficient decorrelation between 
consecutive samplers) and summed to construct the center of mass.  The 
small discrepancy for $N=100$ with the theoretical prediction is due 
to finite size effects.  For $N=1000$, the results are nearly 
indistinguishable on the scale of the figure from the theoretical 
values.  It is interesting to note that the hard constraints of 
eq.~(\ref{c}) are satisfied {\em only\/} in the thermodynamic limit.  
In the simulations $R$ (and $q$) are distributed with peaks whose 
width scales with $N^{-1/2}$.  Nonetheless the effect of these 
fluctuations on the resulting $P(B_{1}Y_{1})$ is negligible.

\stars 

The authors would like to thank the referees for useful suggestions.  
We also acknowledge support from the FWO Vlaanderen and the Belgian 
IUAP program (Prime Minister's Office).


\begin{thebibliography}{99}
\bibitem{ising}
\Name{E. Ising} 
\Review{Z. Phys.}
\Vol{31}
\Year{1925} 
\Page{253}

\bibitem{hertz}
\Name{Hertz A. J., Krogh A. \And Palmer R. G.}
\Book{Introduction to the Theory of Neural Computation}
(Addison-Wesley, Redwood City, CA, USA)
\Year{1991}

\bibitem{mezard} 
\Name{M\'ezard M., Parisi G. \And Virasoro M. A.}
\Book{Spin Glass Theory and Beyond} 
(World Scientific, Singapore)
\Year{1987}

\bibitem{Jaynes} 
\Name{ E. T. Jaynes} in 
\Book{E. T. Jaynes: Papers on Probability, Statistics and 
Statistical Physics} edited by 
\Name{R. D. Rosenkrantz} 
(D. Reidel Publishing Company) 
\Year{1983} 
p\Page{39-76}

\bibitem{Schietse95}
\Name{Schietse J., Bouten M. \And Van den Broeck C.}
\Review{Europhys.  Lett.} 
\Vol{32}
\Year{1995}
\Page{279}

\bibitem{gardner}
\Name{Gardner E. \And Derrida B.}
\Review{J. Phys.  A: Math. Gen.}
\Vol{22} 
\Year{1989}
\Page{1983}

\bibitem{tishby}
\Name{Gy\"orgyi G. \And Tishby N.} in
\Book{Neural Networks and Spin Glasses} edited by  
\Name{W. K. Theumann \And R. K{\"o}berle} 
(World Scientific, Singapore) 
\Year{1990}
p\Page{3-36}

\bibitem{seung}
\Name{Seung H. S., Sompolinsky H. \And Tishby N.}
\Review{Phys.  Rev.  A}
\Vol{45} 
\Year{1992}
\Page{6056}

\bibitem{watkin}
\Name{T. L. H. Watkin}
\Review{Europhys. Lett.}
\Vol{21}
\Year{1993}
\Page{871}

\bibitem{opper} 
\Name{Opper M. \And Haussler D.}
\Review{Phys.  Rev.  Lett.}
\Vol{66}
\Year{1991}
\Page{2677}

\bibitem{gyorgyi}
\Name{G. Gy\"orgyi}
\Review{Phys.  Rev.  A} 
\Vol{41}
\Year{1990}
\Page{7097}

\bibitem{storage}
\Name{Gardner E. \And Derrida B.}
\Review{J. Phys A: Math. Gen.}
\Vol{21}
\Year{1988}
\Page{271}

\end{thebibliography}
\end{document}